\documentclass{article}
\usepackage{arxiv}

\usepackage[utf8]{inputenc} % allow utf-8 input
\usepackage[T1]{fontenc}    % use 8-bit T1 fonts
\usepackage{hyperref}       % hyperlinks
\usepackage{url}            % simple URL typesetting
\usepackage{booktabs}       % professional-quality tables
\usepackage{amsfonts}       % blackboard math symbols
\usepackage{amsmath}       % blackboard math symbols
\usepackage{amssymb}        % blackboard math symbols
\usepackage{nicefrac}       % compact symbols for 1/2, etc.
\usepackage{microtype}      % microtypography
\usepackage{graphicx}
\usepackage{doi}
\usepackage{bm}
\usepackage{ar}

\usepackage{orcidlink}
\DeclareTextFontCommand\textsfi{\usefont{OT1}{cmss}{m}{sl}}
\DeclareMathAlphabet\mathsfi            {OT1}{cmss}{m}{sl}
\DeclareTextFontCommand\textsfb{\usefont{OT1}{cmss}{bx}{n}}
\DeclareMathAlphabet\mathsfb            {OT1}{cmss}{bx}{n}
\DeclareTextFontCommand\textsfbi{\usefont{OT1}{cmss}{m}{sl}}
\DeclareMathAlphabet\mathsfbi            {OT1}{cmss}{m}{sl}
\IfFileExists{t1phv.fd}
{\DeclareTextFontCommand\textsfbi{\usefont{T1}{phv}{b}{it}}
\DeclareMathAlphabet\mathsfbi            {T1}{phv}{b}{it}
}{}
\IfFileExists{ot1phv.fd}
{\DeclareTextFontCommand\textsfbi{\usefont{OT1}{phv}{b}{it}}
\DeclareMathAlphabet\mathsfbi            {OT1}{phv}{b}{it}
}{}

\newcommand{\lapx}{\nabla^2_{x}}

\newcommand{\delx}{\bnabla_{x}}

\newcommand{\pvec}{\mathbf{p}}
\newcommand{\xvec}{\mathbf{x}}

\newcommand{\bnabla}{\boldsymbol{\nabla}}
\newcommand{\bcdot}{\boldsymbol{\cdot}}

\newcommand\Rey{\mbox{\textit{Re}}}  % Reynolds number
\newcommand{\syncc}{~\stackrel{\textstyle \rhd\kern-0.57em\lhd}{\scriptstyle L}~}

\title{Foundation and challenges in modelling Dilute Active Suspensions}

\author{Lloyd Fung \orcidlink{0000-0002-1775-5093} \\
	Department of Applied Mathematics \\ and Theoretical Physics\\
	University of Cambridge\\
	Cambridge, UK, CB3 0WA \\
	\texttt{lsf27@cam.ac.uk} \\
	\And
	{Hakan O. Caldag \orcidlink{0000-0002-4394-6045}} \\
	Department of Mathematics\\
	University of York\\
	York, UK, YO10 5DD \\
	\texttt{hakan.caldag@york.ac.uk} \\
	 \AND
    {Martin A. Bees} \\
	Department of Mathematics\\
	University of York\\
	York, UK, YO10 5DD \\
	\texttt{martin.bees@york.ac.uk} \\
}

\renewcommand{\shorttitle}{Continuum Modelling of Dilute Active Suspensions}

\hypersetup{
pdftitle={Continuum Modelling of Active Suspensions},
pdfsubject={physics.flu-dyn, physics.bio-ph},
pdfauthor={Lloyd Fung},
pdfkeywords={Continuum Model, Active Suspensions, Active Brownian Particles, Microswimmers},
}

\begin{document}
\maketitle

\begin{abstract}
Active suspensions, which consist of suspended self-propelling particles such as swimming microorganisms, often exhibit non-trivial transport properties. 
Continuum models are frequently employed to elucidate phenomena in active suspensions, such as shear trapping of bacteria, bacterial turbulence, and bioconvection patterns in suspensions of algae. 
Yet, these models are often empirically derived and may not always agree with the individual-based description of active particles. 
Here we establish a more rigorous foundation to fully develop a continuum model based on the respective microscopic dynamics through coarse-graining. 
All the assumptions needed to reach popular continuum models from a multi-particle Fokker-Planck equation, which governs the probability of the full configuration space, are explicitly presented. 
In the dilute limit, this approach leads to the mean-field model (a.k.a. Doi-Saintillan-Shelley model), which can be further reduced to a continuum equation for particle density. 
Moreover, we review the limitations and highlight the challenges related to continuum descriptions, including significant issues in implementing physical  boundary conditions and the possible emergence of singular solutions. 
\end{abstract}

% keywords can be removed

\section{Introduction}
Active suspensions are suspensions of self-propelling (or motile) particles whose propelling direction depends on the particles' (stochastic) orientation and their biases to move towards or away from certain external stimuli (called taxes). 
Typical examples include suspensions of motile microorganisms, such as bacteria, spermatozoa, and algae, and artificial microswimmers such as Janus particles.
% Brief showcase of applications of active suspensions
% Active suspensions have gathered much interest across disciplines. 
In soft matter and statistical physics, active suspensions are distinctive in how they are driven out of equilibrium locally when particles expend energy to maintain locomotion.
From a fluid dynamics perspective, they have unusual transport properties due to the particles' motility, and their rich and complex phenomenology in flows with vanishing Reynolds number is nontrivial. The stress they exert on the flow can also give rise to peculiar rheological properties like superfluidity which are not observed in passive fluid systems.
% Application
Understanding the fundamental processes in these complex systems may benefit applications, such as the management of biofilms on surfaces, mixing and unmixing at the microscale, the modelling of phytoplankton and thus the carbon cycle in the ocean, food and beverage production, and the biochemical manufacturing of medicines and biofuels \cite{WANG20101009,mcgillicuddy2019models,bees2014mathematics}.

%History
A key challenge in modelling active suspensions is to derive a macroscopic continuum model that is tractable for analysis and interpretation, while remaining grounded by the essential physics of the microscopic dynamics. 
Early models \cite{Childress1975,Kessler1986} incorporated the effects of taxes by modifying the effective drift in the particle density equation. 
These phenomenological models could only capture certain collective behaviours qualitatively, as they did not take into account the microscopic orientational dynamics.
The seminal work by Pedley \& Kessler \cite{Pedley1990,Pedley1992} made a significant advancement in explicitly accounting for the orientational dynamics within their continuum description.
Taking inspiration from how the mean-field stresses from passive particles are modelled \cite{Hinch1972,Brenner1974}, they wrote down the mean disturbance due to particle motility.
As for the particle transport, they used statistical moments of their orientation distribution to derive the effective drift and diffusivity due to biased motility.
Although the model was based on the microscopic dynamics, the particle's effective diffusivity $D$ remains \emph{ad hoc} and fails to capture the theoretical $D \sim (V^c)^2 / d_r$ scaling \cite{Berg1993}, as Pedley \& Kessler \cite{Pedley1990} separated the orientational and spatial distribution without full consideration of the coupling between the particle's random rotation with rotational diffusivity $d_r$ and its swimming speed $V^c$. 

The more rigorous approach is to model the probability of the particle configuration in both position and orientation space, governed by the Fokker-Planck equation, which is derived from the microscopic Langevin equation.
Hill \& Bees \cite{Hill2002} and Manela \& Frankel \cite{Manela2003} first demonstrated how the coupling between position and orientation space in the Fokker-Planck equation gives rise to an effective dispersion of active particles using the generalised Taylor dispersion theory \cite{Brenner1993}.
Later work by Saintillan \& Shelley \cite{Saintillan2008a} wrote down the same equation and the mean-field flow equation from a kinetic theory perspective.
They used a linear stability analysis to show how the coupling between the particle configuration and the flow caused by particle stresses can lead to collective behaviour at a length scale much larger than the particles themselves. 
Subramanian \& Koch \cite{Subramanian2009} also independently arrived at a similar set of equations for motile bacteria, with an additional term for the run-and-tumble dynamics.
Although the Fokker-Planck approach is more accurate than the previous methods, most of the literature did not explicitly state the model's foundation in the underlying microscopic dynamics and the assumptions used to derive the model.
In particular, few \cite{Baskaran2010} have presented the rigorous reduction from the multi-particle to single-particle Fokker-Planck equation.

% Paragraph about the aim of the article
Therefore, the purpose of this article is to offer a thorough examination of the foundation underlying current continuum models for active suspensions, laying bare all the assumptions necessary to scale up the microscopic dynamics of individual particles to a suspension.
Particular focus will be on the derivation of the mean-field model, sometimes known as the Doi-Saintillan-Shelley (DSS) model, and its limitations due to the dilute assumption.
Other practical issues, such as the high-dimensionality of the model and boundary conditions, will also be discussed.
Finally, we will explore the possibility of adopting coarse-graining methodologies from neighbouring fields to extend continuum modelling to concentrated suspensions.

% Other developments in dense suspensions
Besides rigorous coarse-graining from the bottom up, we should also briefly mention phenomenological approaches to modelling system of active particles. 
Originally developed for flocking, the Toner-Tu theory \cite{Toner1998} proposed that a phenomenological equation can be written down by including all terms allowed by symmetry. 
The theory was extended to suspensions of self-propelled particles \cite{Simha2002} and for modelling active turbulence in dense bacterial suspensions \cite{Wensink2012,Dunkel2013}. 
Another popular approach for modelling active nematic systems, such as a microtubule-kinesin mixtures, is to modify the continuum equations from liquid crystal theory by including new terms arising from the activity \cite{Doostmohammadi2018}.
These models are often associated with very dense systems of active particles, which are beyond the scope of this work.
The review by \cite{Marchetti2013} provides an overview of these phenomenological approaches and their comparisons with coarse-grained models \cite{Baskaran2008,Saintillan2008a}.
Instead, here we shall focus on coarse-graining dilute suspensions of self-propelled particles where hydrodynamic interactions dominate.

% Structure of the article
The article is structured as follows. 
First, we quantify the trajectory and hydrodynamic disturbances arising from a single stochastic self-propelling particle in \S\ref{sec:Microscopics}. 
Then, in \S\ref{sec:NFP-1FP}, we consider $N$ suspended particles and write down the equivalent $N$-particle Fokker-Planck equation and explain the approximation required to reduce it to the mean-field (DSS) model (\S\ref{subsec:MeanField}). 
The resulting one-particle Fokker-Planck equation in the DSS model can be further reduced to a particle density equation (\S\ref{subsec:n-eqn}).
After that, we will discuss several issues resulting from the approximations used to derive the DSS model (\S\ref{sec:challenges}). 
These issues include the emergence of singular solutions due to the lack of volume exclusion in the model (\S\ref{subsec:finite-volume}), difficulties in closing the many-body problem (\S\ref{subsec:BBGKYclosure}), and choice of boundary conditions (\S\ref{subsec:boundary}). 
Finally, in \S\ref{sec:conclusion}, we lay out the direction and priorities in the development of continuum models for active suspensions.

\section{Microscopics: Stokes flow around an active particle}\label{sec:Microscopics}
We start by considering a single rigid active particle suspended in a Newtonian fluid. 
Given the density $\rho_f \sim 1000 \; kg \;  m^{-3}$ and viscosity $\mu \sim 0.001 \textrm{Pa}\; s$ of the fluid , the size of the particle $ a\sim 10-100  \;  \mu m$, and the slip velocity ($\sim 10-100 \; \mu m s^{-1}$), the particle Reynolds number $\Rey_p \approx 10^{-4}- 10^{-2} \ll 1$ is vanishingly small \cite[Fig. 2.2]{Lauga2020}. 
Assuming that the particle $j$ is of volume $\upsilon_j$ and mass $\upsilon_j \rho_p$, 
its inertia can be neglected, as biological active particles usually have density $\rho_p$ close to that of the fluid $\rho_f$.
Hence, the flow around the particle can be treated as steady and Stokesian, governed by
\begin{equation}
    \mathbf{0}= -\mu \lapx \mathbf{u}(\mathbf{x}) +  \delx q(\mathbf{x}) + \mathbf{f}, \qquad \mbox{and} \qquad \delx \bcdot \mathbf{u}(\mathbf{x})=0, \label{eq:Stokes}
\end{equation}
where $\mathbf{u}(\xvec)$ is the flow velocity at location $\xvec$, $q$ the pressure and $\mathbf{f}$ the external forces on the fluid. 
Note that the flow around the particle only truly satisfies the Stokes equation below inertial length scale $\ell_{i} \sim a \Rey_p^{-1}$. The macroscopic flow, however, can be inertial at the phenomenological scale.

\subsection{Disturbance flow due to an active particle}\label{subsec:Disturbance}
First, we consider the disturbance flow around a single particle in an otherwise quiescent flow.
For zero $\Rey_p$, one can exploit the linearity of (\ref{eq:Stokes}) to represent how
a particle centred at $\xvec_j$ exerts stresses from its surface onto the fluid
and creates a disturbance flow velocity $\mathbf{u}^d$. 
At a location $\xvec$ far from $\xvec_j$ (i.e.~at $\mathbf{r}=\xvec-\xvec_j$, $r=|\mathbf{r}| \gg a$) this disturbance flow $\mathbf{u}^d$ can be approximated by the multipole expansions of the Green's function $\mathcal{G}$, such that
\begin{equation}
    \textstyle \mathbf{u}^d(\mathbf{r})= -(8\pi\mu)^{-1} ( {\mathcal{G}}(\mathbf{r}) \bcdot \mathbf{F}_j + \bnabla_x\mathcal{G}(\mathbf{r}) \boldsymbol{:} \mathsf{M}_j +\mathcal{O}(r^{-3}) ), \quad \mbox{where} \quad \mathcal{G}(\mathbf{r})=r^{-1}\mathsf{I} + r^{-3}\mathbf{rr} . \label{eq:Multipole_representation}
\end{equation} % TODO: fix typescript of Identity matrix I
Here, $\mathbf{F}_j$ is the net force and $\mathsf{M}_j$ the net first moment of traction that particle $j$ exerts on the fluid. 
Expansion up to the first-order terms is usually sufficient to represent the flow at large $r$, as terms at order $n$ decay as $1/r^{n+1}$ as $r \rightarrow \infty$; a particle can be approximately represented by a point force and a point dipole in a Stokesian flow.
A point force is called a Stokeslet and is non-zero only when the particle experiences an external net force (e.g.~buoyancy force $\mathbf{F}_j=\upsilon_j \Delta \rho_j \mathbf{g}$ due to gravity $\mathbf{g}$ where $\Delta \rho_j=(\rho_{p,j}-\rho_f)$ is the density mismatch between the particle and the fluid). 
The moment $\mathsf{M}_j$ has an antisymmetric part called a rotlet ($\mathsf{L}_j$, antisymmetric force dipole) and a symmetric part called a stresslet ($\mathsf{S}_j$, symmetric force dipole). These components correspond to the net torque and symmetric first moment of stresses the particle exerts on the flow, respectively.
A net torque $\mathbf{L}_j$ can arise when the centre of hydrodynamic forcing is off-set from the centre of the external force by a finite length $\ell_j$ (e.g.~$\mathbf{L}_j = L_j \mathbf{g} \times \pvec$, $L_j=\ell_j \upsilon_j \Delta \rho_j$ when the particle is bottom-heavy under gravity). 
A stresslet $\mathsf{S}_j$ can arise from the particle's self-propulsion or volume exclusion in the presence of a background flow.
Typically, the stress generated by the self-propulsion of an active particle dominates over other flow-dependent dipoles, and the resulting time-averaged disturbance flow is well-approximated by an axisymmetric force dipole $\mathsf{S}_j=\sigma_j \mathbf{p}_j\mathbf{p}_j$ \cite{Drescher2010a,Drescher2011}, where $\mathbf{p}_j$ is the normalised vector representing the particle's orientation.
A self-propelling particle with $\sigma_j>0$ is called a puller, and $\sigma_j<0$ a pusher. 

\subsection{Typical trajectories of hydrodynamically interacting active particles}\label{subsec:IndDyn}

Now, consider $N$ inertia-free active particles that are far apart from each others in a dilute suspension, i.e. the volume fraction $\phi$ is negligibly small. 
The configuration $\boldsymbol{\xi}_i=(\xvec_i,\pvec_i)$ of each particle $i$ is defined by its position $\mathbf{x}_i$ and orientation $\mathbf{p}_i$.
Collectively, they define the configuration of the suspension $\Xi=(\boldsymbol{\xi}_1,\boldsymbol{\xi}_2,...,\boldsymbol{\xi}_N)$.
Assuming each particle $i$ is axisymmetric along its swimming direction $\pvec_i$,
balancing the external and hydrodynamic forces from self propulsion gives the particle's overdamped trajectory as
\begin{equation}
    \dot{\xvec}(\xvec_i,\pvec_i;\Xi)=\mathbf{u}(\xvec_i,t;\Xi \backslash \boldsymbol{\xi}_i) + V_i^c \pvec_i + \mathbf{V}^F_{i}(\pvec_i). \label{eq:xdot}
\end{equation}
Here, $V_i^c \mathbf{p}_i$ represents the self propulsion, $\mathbf{V}^F_{i}(\pvec_i)$ the orientation-dependent slip velocity due to external force $\mathbf{F}_i$, 
and $\mathbf{u}(\xvec_i,t;\Xi \backslash \boldsymbol{\xi}_i)$ the passive advection by the flow at time $t$, which includes both the background flow and the disturbance from $\Xi\backslash \boldsymbol{\xi}_i$, which is the ensemble of particles other than itself. 
In other words, the trajectory of particle $i$ depends on the configuration of all other particles in the suspension.
We further assume that particles are sufficiently small relative to the flow length scale such that Fax\'{e}n corrections can be omitted (c.f.~\S\ref{subsec:finite-volume}). 

Similarly, the inertia-less torque balance provides the rotational dynamics as
\begin{eqnarray}
	\dot{\pvec}(\xvec_i,\pvec_i;\Xi) &=& (\bm{\Omega}_{\mbox{\mbox{jeff}}}(\mathbf{u}(\xvec_i,t;\Xi \backslash \boldsymbol{\xi}_i),\pvec_i) + \beta_i \pvec_i \times \hat{\mathbf{k}}_i) \times \pvec_i, \label{eq:pdot} \\ 
	\bm{\Omega}_{\mbox{\mbox{jeff}}}(\mathbf{u},\pvec_i) & = & \tfrac{1}{2} \bm{\Omega}+ \alpha_i \pvec_i \bcdot \mathsf{E} \times \pvec_i, \label{eq:Jeffery}
\end{eqnarray} % TODO: Find out the proper typeface for tensor E
where $\bm{\Omega}_{\mbox{\mbox{jeff}}}$ is the Jeffery's torque \cite{Bretherton1962}, $\bm{\Omega}=\bnabla_x \times \mathbf{u}$ the vorticity and $\mathsf{E}=(\bnabla_x \mathbf{u}+\bnabla_x \mathbf{u}^T)/2$ the rate-of-strain tensor.
Jeffery's equation (\ref{eq:Jeffery}) applies to any axisymmetric particle, in which $\alpha_i$ depends on the particle's shape and has the value $(\AR_i^2-1)/(\AR_i^2+1)$ for a spheroidal particle with aspect ratio $\AR_i$.
%%%%%% Gyrotaxis
Meanwhile, the last term of (\ref{eq:pdot}) models any restoring torque arising from bottom-heaviness \cite{Pedley1992} or preferred sedimentation orientation \cite{OMalley2012},
in which $\beta_i$ quantifies the time-scale of the restoration and the unit vector $\hat{\mathbf{k}}_i$ points towards the preferred orientation, which is usually in the opposite direction to $\mathbf{F}_i$.  
As for asymmetric or shape-changing particles, 
readers can refer to a recent review by Ishimoto \cite{Ishimoto2023} on the impact of particle asymmetry on their generalised Jeffery orbits.

Most active particles are large enough that Brownian diffusion due to thermal fluctuation is insignificant. 
However, imperfections or variations in actuation that induce self-propulsion can incur an effective stochasticity to their trajectories. 
For simplicity, we assume that such stochasticity is isotropic and homogeneous white noise in translation and rotation, $\mathbf{W}_x$ and $\mathbf{W}_p$, respectively, where $\mathbf{W}_p$ is much more significant than $\mathbf{W}_x$, except in very close proximity to boundaries (c.f.~\S\ref{subsec:boundary}).
Therefore, one can model an active suspension as a system of $N$ active Brownian particles, in which each particle $i$ has a trajectory given by 
\begin{equation}
	d\mathbf{x}_i(\mathbf{x}_i,\mathbf{p}_i;\Xi) = \dot{\mathbf{x}}(\mathbf{x}_i,\mathbf{p}_i;\Xi) dt + \sqrt{2 D_T} d\mathbf{W}_x ; \quad d\mathbf{p}_i(\mathbf{x}_i,\mathbf{p}_i;\Xi) = \dot{\mathbf{p}}(\mathbf{x}_i,\mathbf{p}_i;\Xi) dt + \sqrt{2 d_r} d\mathbf{W}_p, \label{eq:Stochastic_xp}
\end{equation} 
where $D_T$ and $d_r$ are the translational and rotational diffusivity that model the biological stochasticity, respectively.

While (\ref{eq:xdot}-\ref{eq:Stochastic_xp}) form a complete description of the trajectories of particles $i$ in a given flow field, the flow field also depends on the configuration of other particles. 
More specifically, the flow $\mathbf{u}$ at certain location $\mathbf{x}$ is disturbed by the forcing $\mathbf{F}_j$ and $\mathsf{M}_j$, which are from other particles, and is governed by the incompressibility condition $\bnabla_x \bcdot \mathbf{u}=0$ and the Navier-Stokes equation
\begin{equation}
	\textstyle \rho_f(\partial_t \mathbf{u}+ \mathbf{u}\cdot \bnabla_x \mathbf{u}) = -\bnabla_x q + \mu \nabla_x^2 \mathbf{u} +  \sum_{j} (\mathbf{F}_j+ \bnabla_x \bcdot  \mathsf{M}_j) \delta(\xvec-\xvec_j) - \frac{1}{V}\sum_{j} \mathbf{F}_j. \label{eq:NS-Full} 
\end{equation}
For the flow external to the particles, $\mathbf{u}=\mathbf{u}(\mathbf{x},t;\Xi)$, the summations in (\ref{eq:NS-Full}) are for $j=1,2,...,N$,
whereas the flow experienced by the $i$-th particle $\mathbf{u}(\mathbf{x}_i,t;\Xi \backslash \boldsymbol{\xi}_i)$ will exclude the force on the particle itself ($j=1,2,...,i-1, i+1,...,N$).
Note that the sum of external forces averaged over the domain volume $V$ is subtracted to conserve momentum in the fluid.
\footnote{The removal of the sum of external forces is the equivalent of renormalisation \cite{Hinch1977}, in which a net flow condition is enforced by balancing the total external net forces on the system with the pressure gradient. Under our point particle approximation, the particle has no volume and, therefore, there is no exclusion of regions of overlap that generates extra hindrance to sedimentation. 
Instead, the sum of external forces simply modifies the pressure gradient \cite{Hinch1977}.} 
The full Navier-Stokes equation is invoked here as the macroscopic bulk flow can still be nonlinear and unsteady despite the Stokesian microscopic flow around each particle. Indeed, in some phenomena, such as bioconvection, the nonlinear and unsteady terms play a significant role at the phenomenological scale \cite{Pedley1992,Bees2020}. 
However, coarse-graining the stochastic and nonlinear PDE (\ref{eq:NS-Full}) is very difficult.
Therefore, in the following derivation we shall assume that we have effectively a Stokesian bulk flow.
Linearity and the vanishing $\phi$ approximation allow the superposition $\mathbf{u}(\xvec_i,t;\Xi \backslash \boldsymbol{\xi}_i) = \sum_{j=1,i\neq j}^{N} \mathbf{u}_j(\xvec_i;\boldsymbol{\xi}_j(t))$, where $\mathbf{u}_j$ is the flow due to the presence of each particle $j$, governed by (\ref{eq:Stokes}) with $\mathbf{f}$ substituted by $(\mathbf{F}_j+ \bnabla_x \bcdot  \mathsf{M}_j ) \delta(\xvec-\xvec_j)-\mathbf{F}_j/V$.
Equations (\ref{eq:xdot}-\ref{eq:Stochastic_xp}) and solution to $\mathbf{u}$ form a complete description of the active suspension,
but direct integration of the governing equation for $\mathbf{u}$ is difficult and their results are analytically intractable.
Therefore, coarse-graining is necessary.

\section{Coarse-graining through reducing the Fokker-Planck equation} \label{sec:NFP-1FP}
For the set of stochastic equations (\ref{eq:Stochastic_xp}) for $N$ particles, one can write down the equivalent $N$-particle Fokker-Planck equation, 
 \begin{equation}
	\partial_t \Psi_N(\Xi,t) = \sum_{i=1}^{N} \left\{ \bnabla_{x_i} \cdot \left[ - \dot{\mathbf{x}}(\mathbf{x}_i,\mathbf{p}_i;\Xi)\Psi_N  +  D_T \bnabla_{x_i} \Psi_N \right] + \bnabla_{p_i} \cdot \left[ - \dot{\mathbf{p}}(\mathbf{x}_i,\mathbf{p}_i;\Xi)\Psi_N  +  d_r \bnabla_{p_i} \Psi_N \right] \right\}, \label{eq:FP-Full}
 \end{equation}
governing the joint probability density function (p.d.f.) $\Psi_N$ for the $N$-particle configuration $\Xi$.
The high-dimensional $N$-particle Fokker-Planck equation (\ref{eq:FP-Full}) is intractable without appropriate reduction, 
which we shall demonstrate in the following section, 
but writing (\ref{eq:FP-Full}) down explicitly serves as a reminder that this is the common starting point for most coarse-grained continuum models \cite{Baskaran2010,Bruna2022a,TeVrugt2023}.
For pedagogical simplicity, in this article we will assume all particles are identical, allowing for the removal of the subscript $i$ from particle parameters such as $V^c$, $\mathbf{V}^F$, $\beta$ and $\alpha$.
However, it should be stressed that, in reality, parameters such as $V^c$ can have a large variation among particles, which can significantly impact their collective dynamics \cite{Bees1998}.

\subsection{Averaging towards a one-particle equation} \label{subsec:averagingFP}
Our goal now is to obtain a macroscopic model for the one-particle density by reducing (\ref{eq:FP-Full}) through a series of approximations.
Let us first define the notation for the averaging operator. 
For any function $\star(\Xi)$ of the configuration $\Xi$, the ensemble average over all particle configurations except for the first $k$ particles is defined as
\begin{equation}
	\textstyle \langle \star \rangle_k = \int_{\Upsilon^{N-k}} \star(\Xi) \Psi_N d\boldsymbol{\xi}_{k+1}...d\boldsymbol{\xi}_N ,
\end{equation}
where $\Xi \in \Upsilon^{N-k}$ is the domain of all allowed configurations for the $(k+1)$-th to $N$-th particles. 
The joint p.d.f.~for the first $k$-particles is thus defined by putting $\star=1$, $\Psi_k(\boldsymbol{\xi}_1,...,\boldsymbol{\xi}_k,t) = \langle 1 \rangle_k = \int_{\Upsilon^{N-k}} \Psi_N d\boldsymbol{\xi}_{k+1}...d\boldsymbol{\xi}_N.$
For example, with $\star=1$, putting $k=0$ recovers the normalisation constant $\langle 1 \rangle_0$, which we set to unity, while putting $k=1$ provides the 1-particle p.d.f., 
for which the evolution equation can be obtained by performing the same $\langle \star \rangle_1$ operation on (\ref{eq:FP-Full}), yielding
\begin{eqnarray}
	\partial_t \Psi_1(\xvec_1,\pvec_1,t) & = & \bnabla_{x_1} \cdot \left[ - \langle \mathbf{u} \rangle_1(\mathbf{x_1},t) - \left( V^c \pvec_1 + \mathbf{V}^F(\pvec_1) \right)\Psi_1 +  D_T \bnabla_{x_1} \Psi_1 \right] \nonumber \\
	& + &  \bnabla_{p_1} \cdot \left[-\left(\bm{\Omega}_{\mbox{jeff}}(\langle \mathbf{u} \rangle_1,\pvec_1)  + \beta (\pvec_1 \times \hat{\mathbf{k}})\Psi_1 \right) \times \pvec_1 +  d_r \bnabla_{p_1}  \Psi_1 \right], \label{eq:FP-1}
\end{eqnarray}
where the averaged velocity $\langle \mathbf{u} \rangle_1(\mathbf{x_1,t})$ is defined as
\begin{equation}
	\langle \mathbf{u} \rangle_1(\mathbf{x_1},t)  =  \int_{\Upsilon^{N-1}} \mathbf{u}(\xvec_1,t;\Xi \backslash \boldsymbol{\xi}_1)\Psi_N  d\boldsymbol{\xi}_{2}...d\boldsymbol{\xi}_N  = \int_{\Upsilon^{N-1}} \sum_{j=2}^{N}\mathbf{u}_j(\xvec_1,t;\boldsymbol{\xi}_j)\Psi_N  d\boldsymbol{\xi}_{2}...d\boldsymbol{\xi}_N.\label{eq:uavg-1-multi}
\end{equation}
Since the particles are approximated as flow-independent point forces ($\mathbf{F}_j$ a constant vector) and dipoles ($\mathsf{M}_j=\sigma_j \pvec_j \pvec_j$), there are no three-way particle interactions. 
Therefore, under the assumption that the particles are also indistinguishable, we can further simplify the $N$-particle integral into a $2$-particle integral,
\begin{equation}
	\textstyle \langle \mathbf{u} \rangle_1 (\mathbf{x}_1,t) 
		=  (N-1) \Psi_1(\boldsymbol{\xi}_{1},t) \int_{\Upsilon^1} \mathbf{u}_2(\xvec_1,t;\boldsymbol{\xi}_{2}) \Psi_{2|1} (\boldsymbol{\xi}_{2},t | \boldsymbol{\xi}_{1})  d\boldsymbol{\xi}_{2},  \label{eq:uavg-1-2particle}
\end{equation}
where $\Psi_{2|1} = \Psi_{2} / \Psi_{1}$ is the conditional p.d.f.~of $\xi_2$ given the configuration of the first particle $\xi_1$.
Therefore, defining the average flow (and the corresponding pressure) due to the presence of the other particle as $\tilde{\mathbf{u}}(\xvec,t)=(N-1)\int_{\Upsilon^1} \mathbf{u}_2(\xvec,t;\boldsymbol{\xi}_{2}) \Psi_{2|1} (\boldsymbol{\xi}_{2},t | \boldsymbol{\xi}_{1})d\boldsymbol{\xi}_2$ (and $\tilde{{q}}$), we have 
\begin{eqnarray}
	\partial_t \Psi_1(\xvec_1,\pvec_1,t) & = & \bnabla_{x_1} \cdot \left[ - \left( \tilde{\mathbf{u}}(\xvec_1,t)+ V^c \pvec_1 + \mathbf{V}^F(\pvec_1) \right)\Psi_1 +  D_T \bnabla_{x_1} \Psi_1 \right] \nonumber \\
	& + &  \bnabla_{p_1} \cdot \left[- \left(\bm{\Omega}_{\mbox{jeff}}(\tilde{\mathbf{u}}(\xvec_1,t) ,\pvec_1)  + \beta \pvec_1 \times \hat{\mathbf{k}} \right) \times \pvec_1 \Psi_1 +  d_r \bnabla_{p_1} \Psi_1 \right] \label{eq:FP-1-utilde}
\end{eqnarray}
and
\begin{equation}
	\textstyle \mathbf{0} = -\bnabla_x  \tilde{q}  + \mu \nabla_x^2 \tilde{\mathbf{u}} + (N-1) \left(\int_{p_2}(\mathbf{F}_2 + \bnabla_x \bcdot \mathsf{M}_2)\Psi_{2|1} d\pvec_2 -\frac{\mathbf{F}_2}{V} \right), \quad \bnabla_x \cdot \tilde{\mathbf{u}} = 0. \label{eq:Stokes-utilde}
\end{equation}
Therefore, computing $\Psi_{1}$ requires $\Psi_{2|1} (\boldsymbol{\xi}_2,t|\boldsymbol{\xi}_1)$. This can be found by repeating the above procedures using $\langle \star \rangle_2$ in place of $\langle \star \rangle_1$ to get the $2$-particle equation for $\Psi_{2} (\boldsymbol{\xi}_1,\boldsymbol{\xi}_2,t)$, but the corresponding flow equation will require $\Psi_{3}$, and so on, forming a hierarchy analogous to the BBGKY hierarchy for Hamiltonian systems \cite{Bruna2022a,TeVrugt2023}. 
To close this reduction to a $1$-particle equation requires further assumptions or approximations. In the following section, we will discuss the simplest way to close the problem -- the mean-field approximation. A more general discussion on other potential closure relationships will be discussed at the end of the work.

\subsection{Mean-field approximation} \label{subsec:MeanField}
In the dilute limit $\phi\rightarrow 0$, particles do not interact frequently or strongly, allowing the approximation $\Psi_2(\boldsymbol{\xi}_1,\boldsymbol{\xi}_2,t) \approx \Psi_1(\boldsymbol{\xi}_1,t)\Psi_1(\boldsymbol{\xi}_2,t)$ under the assumption of independence between particle probabilities. With this approximation, we can recover the Doi-Saintillan-Shelley model, which is given by (\ref{eq:FP-1-utilde} - \ref{eq:Stokes-utilde}) with 
$\Psi_{2|1}\approx \Psi_1$ and $\pvec_2 \mapsto \pvec_1$.

At this point, it is a good idea to summarise all the assumptions made. 
(1) Particles are inertialess and small enough to be considered Stokesian. Their overdamped trajectories can be explicitly written out.
(2) Particles are indistinguishable to allow for the simplification from the $N$-particle integral (\ref{eq:uavg-1-multi}) to the $2$-particle integral (\ref{eq:uavg-1-2particle}).
(3) Particle disturbances are represented by the first two terms in the multipole expansion, dependent only on the particle's own configuration and not the others (i.e.~flow-independent); all hydrodynamic interactions are approximated as one-way, which is accurate up to $\mathcal{O}(\phi^{1/3})$ \cite{Kim1991}.
(4) Near-field interactions between particles are neglected; there are no repulsive, lubrication or steric forces between the particles and, therefore, no volume exclusion. 
The whole domain is accessible to each particle. 
(5) There is no correlation between the distribution of any two particles, leading to $\Psi_{2|1}(\boldsymbol{\xi}_2,t|\boldsymbol{\xi}_1) \approx \Psi_1(\boldsymbol{\xi}_2,t)$.
The lack of domain exclusion (4) implies the result is only $\mathcal{O}(1)$ accurate in $\phi$ \cite{Hinch1977}; the DSS model is accurate only when $\phi \rightarrow 0$.
Later in \S\ref{subsec:finite-volume} we will show how the assumption can lead to unphysical artefacts.

Also, it is important to emphasise that this is a mean-field description, which may not always account for all spatial and temporal correlations in the system. 
For example, in bacterial turbulence, even before the onset of instability \cite{Saintillan2008a} where the mean field is isotropic and homogeneous, the statistical correlation between two points can nonetheless indicate the presence of coherent structure \cite{Stenhammar2017}.
In those cases, one may  extend the BBGKY hierarchy to the next order to explicitly calculate the spatial and temporal correlation functions \cite{Skultety2020}.

\subsection{Reducing the high-dimensional Fokker-Planck equation}\label{subsec:n-eqn}
While the DSS model is a good approximation for dilute suspensions and can be rigorously derived from microscopic dynamics, in practice (\ref{eq:FP-1-utilde}) can only be solved numerically in two-dimensional systems \cite{Saintillan2008a}. 
The high dimensionality of the $\xvec-\pvec$ space makes it impossible to solve (\ref{eq:FP-1-utilde}) numerically in three-dimensional systems unless the problem has certain symmetries that lower its dimensionality \cite{Jiang2020,Fung2022,Fung2023a}. 

To overcome this limitation, one approach is to take successive orientational moments of (\ref{eq:FP-1-utilde}), which yield a set of `hydrodynamic' equations for particle density, polar order, and nematic order \cite{Saintillan2005,Baskaran2008} that also resemble the equivalent phenomenological models \cite{Marchetti2013}.
However, this approach gives rise to a hierarchy of equations that require either an \emph{ad hoc} closure \cite{Doi1981,Baskaran2009} or an asymptotic approximation \cite{Hinch1976} of the higher-order moments \cite[\S 2.7.1]{Saintillan2013}.

The second approach involves asymptotically separating the rapid orientational dynamics from the slower translational dynamics, assuming that the orientational distribution relaxes to a quasi-steady state much more quickly than the translational dynamics.
To this end, the generalised Taylor dispersion (GTD) theory has been employed to derive the effective drift and dispersion of tactic particles in simple linear flows \cite{Hill2002, Manela2003}.
However, the particle density equation obtained using GTD in one flow is not intended to be applicable to other flows.
Furthermore, inappropriate application of GTD may result in inaccurate estimations of particle density \cite{Bearon2015} or singularities in the effective diffusivity \cite{Bearon2011}.
To address this issue, Fung \emph{et al.} \cite{Fung2022} proposed the local approximation model.
Similar to Pedley \& Kessler's model \cite{Pedley1992}, this method gives effective drift and dispersion as functions of the local flow field and particle taxes instead of the global flow field like GTD.
It was demonstrated to be accurate in predicting shear trapping phenomena \cite{Rusconi2014} and requires no knowledge of the global flow field, making it more generalisable than the GTD \cite{Bearon2015}.
However, the local approximation model is still limited to cases where orientational dynamics are fast and has limited applicability to fast unsteady flows.

\section{Outstanding challenges in continuum modelling}\label{sec:challenges}
\subsection{Volume exclusion and near-field interactions between particles \label{subsec:finite-volume}}
An underlying assumption in the derivation above is that the particles are treated as point particles, meaning that the volume fraction $\phi \rightarrow 0$. Occasionally, this assumption can result in unphysical outcomes as the model does not account for volume exclusion effects.
To illustrate, let us consider the example of gyrotactic focusing of bottom-heavy ($\beta \propto L > 0$, $\hat{\mathbf{k}} \parallel -\mathbf{g}$), negatively-buoyant ($\Delta \rho >0$) motile microorganisms such as \emph{C. augustae} (n\'{e}e \emph{nivalis}) \cite{Kessler1986,Pedley1992}.
The DSS model was applied to this suspension in a recent work \cite{Fung2023a}, which revealed a surprising equivalence between the DSS model for gyrotactic focussing and the Keller-Segel model for autochemotaxis.
Similar to the well-known chemotactic collapse in the Keller-Segel model \cite{Childress1981}, the steady-state velocity and particle density in the DSS model also tend to infinity locally in three-dimensional systems.
The same singularity was also found when applying the DSS model to the sedimentation of spheroids \cite{Fung2023a} and chemotactically active immotile particles on gas-fluid interfaces \cite{Masoud2014}.
Although the aggregation of particles does corroborate experimental observations, the unrealistic blow-up of local particle density shows the importance of incorporating the volume exclusion effect.

The finite volume of particles can necessitate multiple corrections:
(1) The passive advection $\mathbf{u}(x_i,t,\Xi \backslash \boldsymbol{\xi}_i)$ should be corrected using the Fax\'{e}n law to account for variation in the flow at the length scale of the particle. 
(2a) The naive superposition of multipole expansions of Green's function as a representation of disturbance by each particle is no longer strictly valid, but remains a good approximation by the method of reflections. 
The finite size of particles implies that each particle will exert an extra stresslet depending on the local strain rate due to background flow and disturbance flow from other particles. %This is an $\mathcal{O}(\phi)$ effect.
(2b) Since this stresslet depends on the configuration of other particles, three or more particle interactions become finite (but small),
and the step from (\ref{eq:uavg-1-multi}) to (\ref{eq:uavg-1-2particle}) is no longer exact. 
(3) Volume-exclusion implies that the domain of $\Psi_{2|1}$ has impenetrable regions, which can induce significant corrections. For example, in the sedimentation problem \cite{Hinch1977}, it is the exclusion of this region during renormalisation that induces a correction to the average sedimentation speed, which is $-6.55\phi$ including the Fax\'{e}n correction. 
(4) There is a non-zero chance that particles will come into close contact with each other. When their surfaces are at a distance $h \lesssim a$, the far-field multipole expansion of the disturbance flow is no longer valid, and the flow field around two particles must be calculated explicitly \cite{Jeffrey1984,Papavassiliou2017}. 
(5) As a result of the near-field interactions, the pairwise correlation $\Psi_{2|1}$ is likely not well approximated by $\Psi_{2|1} \approx \Psi_1(\boldsymbol{\xi}_2,t)$, and the closure relationship between $\Psi_{2|1}$ and $\Psi_1$ near particle $1$ must be re-examined.

Methodologies to account for (1-3) are well-established \cite{Hinch1977} if the pair correlation $\Psi_{2|1}$ is known,
but the exact physics of the near-field interactions (4) is still an active area of research even in passive suspensions \cite{How2021} due to lubrication breakdown \cite{Ball1995}.
Moreover, past work on passive suspensions showed that the approximation of $\Psi_{2|1}$ in (5) is highly sensitive to the exact near-field interactions in some problems due to the singular nature of pairwise trajectories in Stokes flow \cite{Brady1997}. 
As a result, macroscopic properties like the bulk stress in the sheared passive suspension are hard to predict, and are highly sensitive to the near-field interactions governed by the geometry and surface properties of the particles \cite{Brady1997}. 
For self-propelling particles such as swimming microorganisms and artificial microswimmers, deriving macroscopic models that account for the finite volume will be more challenging as the near-field interactions will likely depend on the specific propulsion mechanics and microscale flow around the particles.
While exact analytical results are available for simple spherical squirmers \cite{Papavassiliou2017}, there is no general theory to calculate near-field interactions between arbitrary particles other than employing prohibitive numerical methods. 
Moreover, it is expected that real biological microswimmers will have a wide variety of shapes and sizes, making the calculation of near-field interactions intractable.
Therefore, it remains an open challenge to derive a general framework that can include complex and wide variations of two-particle interactions in a coarse-grained model.

\subsection{Overcoming the closure problem in many-body systems} \label{subsec:BBGKYclosure}
The above discussion focuses on pairwise interactions, which, when resolved, can improve the model accuracy at finite volume fractions. 
To extend the approximation to higher $\phi$, many-body interactions should be considered \cite{Brady1997}.
However, as demonstrated by the derivation of (\ref{eq:Stokes-utilde}), the consideration of $N$-body interactions in equation for $\Psi_{N}$ always require information from $\Psi_{N+1}$. 
This is the BBGKY hierarchy, which is a fundamental challenge in many-body systems.
While there are many approaches to overcome this closure problem, here we present two recent developments in neighbouring fields that may offer new opportunities in addressing the problem in the context of active suspensions.
Reviews by \cite{TeVrugt2020,Bruna2022a,TeVrugt2023} offer an introduction to alternative approaches.

\paragraph*{Dynamical Density Functional Theory}
Density functional theory (DFT) is a powerful tool in statistical mechanics that is used to write down the equilibrium description of an interacting particle system as a one-particle density that minimises the free energy functional.
It determines the distribution correlation between particles by using an equilibrium argument, effectively closing the BBGKY hierarchy. 
In dynamical density functional theory (DDFT), the idea of solving distribution correlation using free energy functionals is extended to dynamically evolving systems, where the pairwise correlation in the dynamically evolving non-equilibrium system is approximated to be the same as that of the equilibrium system with the same one-body density.
This heuristic approximation is known as the adiabatic approximation. 
DDFT has been used to study the dynamics of hydrodynamically interacting particles \cite{Goddard2012}, `dry' active Brownian particles \cite{Wensink2008}, and, most notably here, for microswimmers (i.e., suspended active particles) \cite{Menzel2016}.
A caveat of the theory, however, is that for the adiabatic approximation to remain valid, the dynamics of the system must evolve close to equilibrium, which is particularly problematic for active particles where the system is driven far from equilibrium by the activity of the particles.
Therefore, the approach may be limited to systems with weak activity.
The review by \cite{TeVrugt2020} provides a good summary of the method, its limitations, and a comprehensive list of applications of theory in various fields.

\paragraph*{Matched asymptotic method for excluded-volume active particles}
Bruna \& Chapman \cite{Bruna2012} demonstrated a novel matched asymptotic technique to account for the excluded-volume effect when reducing the $N$-particle Fokker-Planck equation into a one-particle equation for a system of interacting hard-sphere particles. 
Focusing on the interaction between two spheres, Bruna \& Chapman separated the inner region near the first sphere, where the probabilities of two spheres are correlated, from the outer region, where the two spheres are uncorrelated, via asymptotic expansion in the limit of small volume fractions. 
Then, by matching the inner and outer solutions, Bruna derived a one-particle equation that corrects the drift and diffusion terms to first order in $\phi$.
Recently, Bruna extended the method to active Brownian particles \cite{Bruna2022a} under the assumption that the particles only interact through a hard-core potential.
It remains to be seen if the method can be extended to suspended active particles, where both long-range hydrodynamic interactions and short-range lubrication forces are present. 
A challenge in extending the method to suspended particles is that hydrodynamic interactions are long-range, in contrast to the short-range hard-core potential of previous work. 

\subsection{Prescribing physically relevant boundary conditions}\label{subsec:boundary}
A further challenge in developing continuum models for active particles is prescribing a mathematically well-posed and physically correct boundary condition that captures the interaction between the particles and the boundary.
Active particles are well known to exhibit unique interactions with the boundary. Observations include bacteria swimming in circles \cite{Lauga2006}, attraction to or repulsion from the boundary \cite{Berke2008}, upstream swimming near boundaries \cite{hill2007hydrodynamic,rusconi_microbes_2015} and biofilm formation \cite{sauer2022biofilm}. These phenomena are primarily related to the existence of a plane boundary. Observations diversify even further if one considers constrictions \cite{maleki_activity-induced_2024,ostapenko_curvature-guided_2018}, the geometric complexity of the boundaries \cite{Chen2021} and/or behavioural change of the particle \cite{Zeng2022}. 
Several reviews are available summarising such findings \cite{aranson_bacterial_2022,lauga_bacterial_2016,Elgeti2015}. 
\subsubsection{Individual interactions with the boundary}
%% Hydrodynamic interactions
% Individual-level, Lauga's equations
The existence of a boundary brings three kinds of interactions into consideration: hydrodynamic, steric and behavioural. 
We denote contact-free interactions via the fluid medium as hydrodynamic, the contact-driven interactions as steric, and any biological response near the boundary as behavioural.
Much focus has been given to the debate on whether hydrodynamic or steric interactions dominate the entrapment of pusher particles such as bacteria near the boundary \cite{Berke2008,Li2009,Spagnolie2012,Costanzo2012,Elgeti2013a,Ezhilan2015}, as both hydrodynamic and steric interactions can give rise to accumulation of particles near the boundary. Recent experiments \cite{Drescher2011,Kantsler2013,Molaei2014,Contino2015,Bianchi2017} have shown a more nuanced picture where, depending on the exact geometry and swimming mechanisms of the particle, both hydrodynamic and steric interactions can play a role in the entrapment or scattering of particles at the boundary.
However, one should also note the potential for behavioural changes of microorganisms near the boundary (e.g. \cite{Zeng2022}), which can become dominant and further complicate the picture.
Here, we shall not discuss how these mechanisms give rise to different phenomenology, but merely highlight how one can account for these effects in a continuum model.

Individually, hydrodynamic interactions with the boundary can be represented as corrections to the translational and rotational dynamics that depend on the position and orientation relative to the boundary.
For example, a pusher ($\sigma>0$) is attracted to a no-slip wall and will reorient to swim parallel to the wall due to the reflected disturbance flow from the axisymmetric force dipole $\mathsf{S}=\sigma \mathbf{p}\mathbf{p}$ \cite{Berke2008}.
Surprisingly, the far-field representation of the particle (\ref{eq:Multipole_representation}) is accurate enough to capture the essential boundary interactions even when the distance between particle and boundary is a fraction of the particle length scale \cite{Spagnolie2012}.
Further corrections can be built upon this framework, such as time dependency due to flagella undulation \cite{Walker2023} and noise-induced drift due to wall-induced spatial gradient of particle motility \cite{Leishangthem2024}.
We refer the reader to \cite[Ch. 11]{Lauga2020} for more details on hydrodynamic interactions at the boundaries.

Regardless of complexity, the hydrodynamic interactions with the boundary can always be represented by a correction to (\ref{eq:xdot} - \ref{eq:Jeffery}) if particles are assumed to only affect themselves. 
Alternatively, extra disturbance in the form of (\ref{eq:Multipole_representation}) from the image of the particle can be added to the coarse-graining framework in previous sections if hydrodynamic interactions between particles via reflection of the wall is of concern.
Either way, one can incorporate hydrodynamic interactions with the boundary in a mean-field continuum model simply by adding the correction terms in either (\ref{eq:Multipole_representation}) or (\ref{eq:xdot} - \ref{eq:Jeffery}), and follow the coarse-graining procedure laid out in the previous section. Nonetheless, continuum modelling of hydrodynamically interacting particles near the wall remains sporadic \cite{Schaar2015}.

% Steric interactions
As for steric interactions, on an individual level, the common approach is to empirically add a repulsive force from a wall potential \cite{Hernandez-Ortiz2009,Costanzo2012,Sepulveda2018}.
While some authors have opted for smooth approximations to potential functions that are easier to treat analytically \cite{Drescher2011}, the physical origin of such contact force merits more attention \cite{Poortinga2002}. 
Although the reorientation of active particles near walls is often attributed to hydrodynamics, it should be noted that a contact force from the boundary can also cause particle reorientation if the particle is not spherical and the force acts off-centre \cite{Kantsler2013,Bianchi2017}.
Much like hydrodynamic interactions, the effects of repulsive forces can be easily accounted for by adding correction terms to (\ref{eq:xdot} - \ref{eq:Jeffery}).
\subsubsection{Boundary conditions for the Fokker-Planck equation at different scales}
% Robin boundary condition
When particles are stochastic, volume exclusion further restrains the stochastic force from pushing the particles through the boundary, which requires further treatment in both individual-based and continuum models.
At the individual and microscopic level, the conventional method to prevent particles experiencing thermal Brownian motion from crossing the boundary is to reflect their positions about the boundary but not their orientations \cite{Elgeti2009,Schuss2013}
\footnote{The word "reflecting" boundary is taken from the convention \cite{Schuss2013} on modelling random walks close to a boundary, where a random walker is reflected about the boundary when it crosses, while keeping the configuration in the dimensions orthogonal to the wall direction unchanged. To avoid confusion with boundaries that reflect both the position and orientation of the particle like a pin-ball, we will use the term "specularly reflecting" boundary to refer to the latter case where both $\pvec$ and position in the wall-normal direction are reflected.}. 
This reflection ensures that no particle flux passes through the boundary.
At the continuum level, this treatment is equivalent to the no-flux Robin boundary condition \cite{Elgeti2013a,Ezhilan2015,Jiang2019}, given by
\begin{equation}\label{eq:robin_ezhilan}
    D_T \frac{\partial \Psi}{\partial  \mathbf{n}} = V_s \mathbf{p} \cdot \mathbf{n} \Psi,
\end{equation}
where $\mathbf{n}$ is the normal vector to the boundary.
Since the balance between the swimming and diffusive flux is enforced at each $\pvec$, (\ref{eq:robin_ezhilan}) will lead to a boundary layer of polarised particles accumulating near the wall with a thickness scaling with $D_T$.
Some authors \cite{Elgeti2013a,Elgeti2016,Ezhilan2015} argue that the near-wall accumulation is in qualitative agreement with experimental observations \cite{Rothschild1963,Li2009}, although further work is required to establish this conclusively.
Others interpret the boundary condition as unrealistic \cite{maretvadakethope_interplay_2023} and over-restrictive, as (\ref{eq:robin_ezhilan}) prescribes that particles do not reorient upon contacting the wall. 
Also, for many realistic active particles, $D_T$ is vanishingly small, which may lead to an impractical boundary layer thickness or ill condition in (\ref{eq:robin_ezhilan}).
Here, it is important to note that $D_T$ does not represent the thermal diffusivity because active particles of $\mu m$ size are too large to experience significant translational Brownian motion (c.f. \S\ref{subsec:IndDyn}). 
Instead, it is exploited as a 'catch-all' model for any physical, stochastic or biological behaviour that is not captured in (\ref{eq:xdot}-\ref{eq:Jeffery}). 
The question of how these behaviours change upon contact with the boundary remains unanswered, but it is probable that simple reflection off the wall without reorientation would be an oversimplification.
Nonetheless, in theoretical studies of reorientational dynamics in stochastic point/spherical particles near a boundary due to hydrodynamic and contact forces, it is still reasonable to use the reflection or the corresponding Robin boundary condition (\ref{eq:robin_ezhilan}) to keep particles from crossing the boundary, as they are mathematically consistent with the bulk and do not introduce any additional reorientation dynamics.
As for non-spherical particles, the admissible position of the centre of the particle constrained by the boundary depends on the particle shape and $\pvec$, thereby creating a complex boundary geometry in the $\mathbf{x}-\pvec$ space. 
To this end, Chen \& Thiffeault \cite{Chen2021} proposed a mapping method to specify the boundary geometry according to the particle shape.

While the above boundary conditions hold theoretical significance for studying the microscopic dynamics near the wall, their applications in confined macroscopic systems are often impractical due to the large separation of scales between the microscopic dynamics near the wall and the bulk scale dynamics.
Moreover, modelling realistic biological swimmers with the above methods often proves too complex.
For example, real biological swimmers have complex time-varying geometry, like beating cilia or flagella, to generate propulsion, deeming Chen \& Thiffeault's method challenging to apply.
In other cases, behavioural changes near the boundary, such as the sharp turn of algae \cite{Zeng2022} and the change in phenotype of biofilm forming bacteria \cite{sauer2022biofilm}, dominate over other physical interactions.

Considering the challenges to accurately model boundary interactions, a more feasible way to represent them at the bulk scale may be to encompass all of the above effects in an empirical boundary condition inferred from statistical data of particle scattering or entrapment at the boundary \cite{Spagnolie2012, Kantsler2013, Contino2015}.
For example, the sharp turnaround of algae \cite{Kantsler2013,Zeng2022} motivates a specular reflection boundary condition \cite{Bearon2011,Volpe2014},
\begin{equation}\label{eq:specular}
    \Psi(\mathbf{x},\mathbf{p},t) = \Psi(\mathbf{x},\mathbf{p}+2(\mathbf{p}\cdot \mathbf{n})\mathbf{n},t),
\end{equation}
where particles bounce off the wall like a billiard ball with the angle of reflection equal to the angle of incidence.
This condition is popular among the recent work in continuum models \cite{Jiang2019,Jiang2020,wang2022transient} due to its ease of implementation and admission to simple solutions such as a uniform steady solution \cite{Ezhilan2015}, but it is likely an oversimplification of the real scattering correlation between incoming and outgoing angles, which often shows particles emerging at a fixed angle irrespective of the incoming angle \cite{Kantsler2013,Spagnolie2012}.
In other cases, particles may emerge from the wall with a wide distribution of orientation \cite{Contino2015}, which motivates a uniform random reflection condition in the limiting case.
The entrapment of some bacteria on the surface and their eventual formation of biofilms may motivate an absorbing boundary condition.
The effects of these possible limiting boundary conditions on the bulk were explored in \cite{maretvadakethope_interplay_2023} in the context of a channel flow where the absorbing boundary is simply represented by $\Psi=0$ on the boundaries for all orientations.

% On the particle density level.
\subsubsection{Boundary conditions for the particle density equation}
As mentioned in \S\ref{subsec:n-eqn}, the equation for the probability density $\Psi(\xvec,\pvec,t)$ in $\xvec$ and $\pvec$ can be further reduced into an equation for the particle density $n(\xvec,t)$.
The most basic boundary condition to apply in this case is the no-flux condition, 
where the advective flux due to motility must balance the diffusive flux, such that
\begin{equation}
    \mathbf{n}\cdot (\bar{\mathbf{V}} n - \bar{\bm{D}} \cdot \delx n)=0 \quad \text{on} \quad \Omega.
\end{equation}
Here, $\bar{\mathbf{V}} $ and $\bar{\bm{D}}$ can be modelled using the effective drift and diffusivity derived from the reduction in the bulk (\S\ref{subsec:n-eqn}),
but using the bulk values would be an oversimplification, as particle trajectories are modified by boundaries. 
However, given the aforementioned difficulties in accounting for the exact interactions between the particles and the boundary, there is little work on how $\bar{\mathbf{V}} $ and $\bar{\bm{D}}$ should be modified by the boundaries.

Beside the no-flux boundary, it is also possible to have an absorbing boundary at the macroscopic scale, given particles can be trapped at the boundary at the microscopic scale physically or biologically.
The condition has practical relevance in terms of biofilm formation, a phenomenon detrimental to many systems such as bioreactors and catheters \cite{zeriouh2017biofouling}. This will require the boundary conditions to be updated to the form
\begin{equation}
    \mathbf{n}\cdot (\bar{\mathbf{V}}  n - \bar{\bm{D}} \cdot \delx n)=-\gamma n \quad \text{on} \quad \Omega,
\end{equation}
where $\gamma$ appears as an absorption coefficient. This condition renders the classical methods founded upon no-flux boundaries cumbersome to manipulate \cite{sankarasubramanian1973unsteady}, requiring different approaches \cite{shapiro1987chemically,lin_taylor_2019,wang2022transient}.

\section{Conclusion and outlook}\label{sec:conclusion}
% Summary of limitation of the mean-field model
In this work, we have presented the systematic derivation of a mean-field model for dilute active suspensions from the microscopic dynamics of active particles and their far-field hydrodynamic interactions.
In particular, we have clarified the approximations required to reduce the $N$-particle Fokker-Planck equation (\ref{eq:FP-Full}) to the one-particle mean-field equation (the DSS model). 
The approximations assume that suspensions are dilute ($\phi \rightarrow 0$) and the probability of each particle configuration is independent of the others ($\Psi_{2|1} \approx \Psi_{1}$). 
However, as the mean-field model does not account for volume exclusion between particles it can lead to unphysical artefacts, such as particle density singularities when modelling gyrotactic focusing \cite{Fung2023a}.
Moreover, recent work on bacterial turbulence has shown that the mean-field model is not sufficient for describing the full dynamics of the system \cite{Stenhammar2017}.
Even when the mean field is isotropic and homogeneous, fluctuations may be correlated, leading to the emergence of bacterial turbulence  \cite{Skultety2020}.

% Reduction of FP equation 
We also briefly reviewed the approximation techniques for further reducing the high-dimensional one-particle Fokker-Planck equation.
One strategy, popular among soft matter physicists, led to the `hydrodynamic' equations for self-propelling particles \cite{Saintillan2005,Baskaran2008}.
Others provide an effective transport equation for particle density, assuming that the orientational distribution rapidly converges to a quasi-steady equilibrium \cite{Hill2002,Manela2003,Fung2022}.
However, one should exercise caution when interpreting the resulting transport model, as some methods, such as the generalised Taylor dispersion model, are based on the restrictive assumption of homogeneity in the global flow field.
Fortunately, in the advent of increasing computational power, direct numerical simulation of the Fokker-Planck equation may soon become practical, with recent methods \cite{Jiang2019,Jiang2020,Fung2022} tackling four of the five dimensions in the equation.

% Key Challenge 1: Near-field interactions
To further develop the coarse-graining framework into a mature and quantitatively accurate model for active suspensions, there remain several major challenges.
Firstly, recent results on phenomena such as gyrotactic focusing suggest that one should account for the finite volume of the particles and their near-field interactions \cite{Fung2023a}.
Although some analytical treatments for near-field interactions between active particles in simple geometries exist \cite{Papavassiliou2017}, most studies on such interactions rely on well-established numerical methodologies (e.g.~\cite{Ishikawa2006}). 
There is also a trend towards higher fidelity in near-field models, where particle geometries are increasingly complex and the flow increasingly intractable.
Therefore, developing coarse-graining frameworks that can incorporate the numerical results of near-field interactions is a pressing challenge.

% Key Challenge 2: Closure problem in many-body systems
Secondly, there needs to be more development beyond the current mean-field model, which the field has relied on since early 2000s \cite{Hill2002,Manela2003}.
To include the volume exclusion effect and better account for two-particle correlations, one could borrow techniques from the classical literature \cite{Brady1997,Hinch1977} on passive suspensions and extend them to active suspensions.
Further improvements may be made by considering three body interactions and beyond using the methods in \S\ref{subsec:BBGKYclosure}. 
Alternatively, the neighbouring fields of plasma physics and self-gravitating matter may also offer valuable lessons in modelling many-body systems with long-range interactions.

% Key Challenge 3: boundary conditions
Finally, modelling particles close to boundaries remains a significant challenge.
The physical interactions between individual particles and the boundary have received a lot of attention \cite{Li2009,Spagnolie2012,Costanzo2012,Elgeti2013a,Ezhilan2015,Drescher2011,Kantsler2013,Molaei2014,Contino2015,Bianchi2017}, but the continuum modelling of these interactions is sparse.
As the interactions occur at the particle length scale, the particle density can form a microscale boundary layer. To simplify computations, it might be necessary to further homogenise the boundary interactions into an effective boundary condition at the macroscopic scale.
Also, as biological microswimmers can alter their behaviour close to the boundary \cite{Zeng2022}, it may be more practical to adopt a data-centric approach to the creation of macroscopic effective boundary conditions. 
Rather than establishing the boundary interactions from physical principles, statistical data on particle scattering or entrapment at the boundary \cite{Spagnolie2012, Kantsler2013, Contino2015} can be used to infer empirical boundary conditions for macroscopic continuum models.

% Outlook
Aside from these key challenges, there are other important outstanding questions.
For example, biological swimmers are typically propelled by relatively long rotating or beating flagella, but these may interact hydrodynamically and synchronise when in close proximity \cite{Schoeller2018}. 
How does one account for the effect of synchronisation and indeed the complex flagellum-wall interactions in a continuum model?
In addition to the noise in orientation $d_r$, bacteria also tumble randomly. How does run-and-tumble motion, or more generally taxes, change near the boundary and how does one coarse-grain such change in behaviour?
In realistic microswimmer suspensions, swimming speed and gyrotactic strength can vary widely in the population. How does polydispersity influence their collective dynamics \cite{Bees2020}? 
These questions present additional challenges to the coarse-graining framework described in this work.

\paragraph*{Acknowledgements}
We gratefully acknowledge E. J. Hinch for his guidance on coarse-graining, P. H. Htet for his assistance in preparing the discussion on wall interactions, E. Yeo, W. Ridgway, W. Liao and M. te Vrugt for their helpful feedback on the manuscript. 
L.F. is supported by the Research Fellowship from Peterhouse, Cambridge.

\bibliographystyle{RS}
\bibliography{FullLib,HakanLib} 

\end{document}